\documentclass[aip,apl,preprint,superscriptaddress, showpacs, showkeys]{revtex4-1}
\usepackage{graphicx}
\usepackage{dcolumn}
\usepackage{bm}
\usepackage{amsmath}
\DeclareMathOperator{\sign}{sign}

\begin{document}

\title{Detection of shorter-than-skin-depth acoustic pulses in a metal film via transient reflectivity}
\date{\today}
\author{K. J. Manke}
\email[Author to
whom correspondence should be addressed. Electronic mail:
]{kjmanke@mit.edu}
\affiliation{Department of Chemistry, Massachusetts Institute of
Technology, Cambridge, Massachusetts, 02139, USA}
\author{A. A. Maznev}
\affiliation{Department of Chemistry, Massachusetts Institute of
Technology, Cambridge, Massachusetts, 02139, USA}
\author{C. Klieber}
\affiliation{Department of Chemistry, Massachusetts Institute of
Technology, Cambridge, Massachusetts, 02139, USA}
\author{V. Shalagatskyi}
\affiliation{Institut des Mol\'ecules et Mat\'eriaux du Mans, UMR
CNRS 6283, Universit\'e du Maine, 72085 Le Mans cedex, France}
\author{V. V. Temnov}
\affiliation{Institut des Mol\'ecules et Mat\'eriaux du Mans, UMR
CNRS 6283, Universit\'e du Maine, 72085 Le Mans cedex, France}
\author{D. Makarov}
\affiliation{Institute for Integrative Nanosciences, IFW Dresden, 01069 Dresden, Germany}
\author{S.-H. Baek}
\affiliation{Department of Materials Science and Engineering,
University of Wisconsin-Madison, Madison, Wisconsin, 53706, USA}
\author{C.-B. Eom}
\affiliation{Department of Materials Science and Engineering,
University of Wisconsin-Madison, Madison, Wisconsin, 53706, USA}
\author{K. A. Nelson}
\affiliation{Department of Chemistry, Massachusetts Institute of
Technology, Cambridge, Massachusetts, 02139, USA}

\begin{abstract}
The detection of ultrashort laser-generated acoustic pulses at a metal
surface and the reconstruction of the acoustic strain profile are investigated. A 2 ps-long acoustic pulse generated in an SrRuO$_{3}$ layer
propagates through an adjacent gold layer and is detected at its
surface by a reflected probe pulse. We show that the intricate
shape of the transient reflectivity waveform and the ability to
resolve acoustic pulses shorter than the optical skin depth are
controlled by a single parameter, which is determined by the ratio of the
real and imaginary parts of the photoelastic constant of the
material. We also demonstrate a Fourier transform-based algorithm that can be used to extract acoustic strain profiles from transient reflectivity
measurements.
\end{abstract}

\pacs{78.20.hc, 68.60.Bs, 68.55.} \keywords{Laser Ultrasonics,
Thin Film Morphology}

\maketitle

Laser-based techniques for the generation and detection of short
acoustic pulses have enabled ultrasonic measurements at the
picosecond time scale and at frequencies up to and beyond 1~THz
\cite{Grahn_IEEE_1989-1, Yamamoto_PRL_94-1, Sun_PRL_00-1, Maznev_Ultrasonics_2012-1}. The optical detection of ultrashort acoustic signals is
often performed by measuring the transient reflectivity of thin
metal films \cite{Thomsen_PRB_86-1}. Because the probe light
penetrates into the material over a finite distance given by the
optical skin depth of the metal, the transient reflectivity signal
does not directly reproduce the strain profile in the acoustic
pulse \cite{Thomsen_PRB_86-1,Devos_PRL_01-1}. Recently Mante et
al.\cite{Mante_PRB_10-1} demonstrated the detection of acoustic
pulses significantly shorter than the optical absorption depth in
semiconductor InP. The question remained whether
shorter-than-skin-depth pulses could be measured with commonly
used metal films, and, more importantly, whether the acoustic
strain profile could be extracted from the intricate waveforms
observed by Mante et al. Most recently, time-resolved
plasmonic interferometry demonstrated that it is possible to measure
acoustic pulses shorter than the skin depth of a surface
plasmon in gold \cite{Temnov_NPhoton_12-1,Temnov_NatComm_13-1}. It
would be interesting to explore the capability of much simpler
transient reflectivity measurements for the quantitative
reconstruction of acoustic pulses.

In this paper, we investigate the transient reflectivity detection
of ultrashort acoustic pulses in gold. We show that the
reflectivity response of any strongly absorbing material to a
shorter-than-skin-depth acoustic pulse can be decomposed into two
terms. The first term contains a sharp step caused by the change
in sign during the acoustic reflection at the free interface. This
sharp step gives access to the acoustic pulse shape with
sub-skin-depth spatial resolution, whereas the second slowly
varying term is not sensitive to the structure of ultrashort
acoustic waveforms. We identify the combination of photoelastic
and optical constants that controls the ability to measure
shorter-than-skin-depth pulses. Furthermore, we demonstrate a
pulse reconstruction algorithm that can be used to recover the
acoustic pulse profile from transient reflectivity measurements.

\begin{figure}
\includegraphics[width = 8cm]{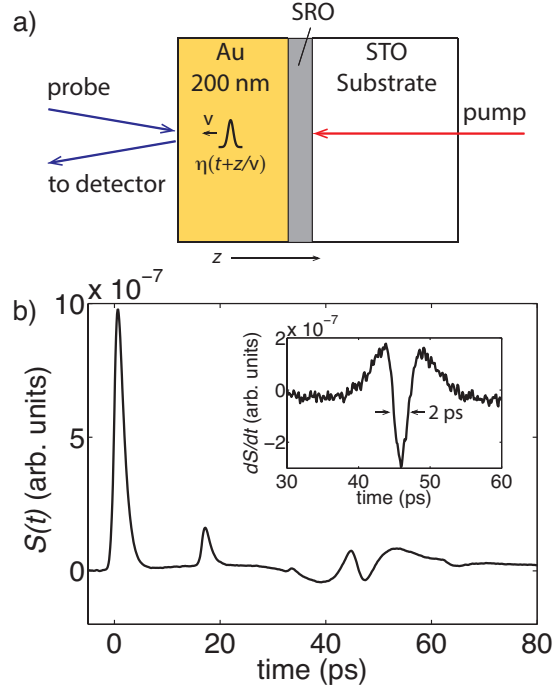}
\caption{\label{fig1}(a) Schematic diagram of the sample structure
and the experimental geometry; (b) the transient reflectivity signal
$S(t)$ detected at the free surface of the gold layer shows the
arrival of the acoustic pulse at a delay time of 46~ps. The inset
shows the time derivative $dS(t)/dt$ of the reflectivity signal.}
\end{figure}

We use a gold/air interface for the detection of acoustic pulses
generated in a very thin layer of single crystal SrRuO$_{3}$ (SRO) sandwiched
between the gold layer and an SrTiO$_{3}$ (STO) substrate. The experimental
geometry for the transient reflectivity measurements is pictured in
Fig.~\ref{fig1}(a). A single crystal SRO layer\cite{Eom_Science_92-1} with a thickness of 12~nm was
grown using 90 degree off-axis RF magnetron sputtering\cite{Eom_APL_89-1} at a rate
of 1 nm/min. A polycrystalline gold film with a thickness of ~200 nm and a primarily (111) texture, as determined by electron
backscattering diffraction, was grown
using magnetron sputtering. High-resolution scanning electron
microscope (SEM) images of the gold film indicated that it had an
average grain diameter of $\sim 400$~nm.

A Ti-sapphire oscillator with a regenerative amplifier was
employed to generate laser pulses with a central wavelength of
786~nm and duration of 300~fs. The laser output was split into
separate pump and probe beams and the probe beam was
frequency-doubled to a wavelength of 393~nm. Illumination of the
SRO transducer layer with the pump pulse through the 1~mm thick
STO substrate launched an acoustic strain pulse through the gold
layer, which was detected at the gold/air interface through the
transient reflectivity of the variably delayed probe pulse.

The reflectivity signal $S(t)$ is depicted in Fig.~\ref{fig1}(b).
The sharp spike at zero pump-probe delay time arises from the
optical generation of hot electrons at the gold/SRO interface
followed by their ultrafast diffusion through the 150~nm thin gold
layer on a sub-picosecond time scale \cite{Brorson_PRL_87-1}. The barely visible
slow rise in the background signal is due to the subsequent thermal
diffusion \cite{Tas_PRB_94-1,Temnov_NatComm_13-1,Lejman_ArXiv_13-1}.
The small subsequent peaks at delay times of 18~ps and 35~ps are
caused by multiple roundtrips of the residual pump reflections
within the 1~mm thick (STO) substrate. The optical signal at 46~ps
marks the arrival of the acoustic strain pulse at the gold/air
interface after propagating through the layer of (111) gold at the
speed of sound $v=3.45$~nm/ps \cite{Anderson_65-1}. It contains a
fast 2-ps transient on top of a slowly varying component. The
duration of the transient corresponds to the expected acoustic
pulse duration generated in the SRO film given the speed of sound
in SRO 6.31 m/s\cite{Yamanaka_JSSC_04-1}.  In gold, this duration corresponds to a
distance of 7 nm, significantly shorter that the skin depth of
16.5 nm\cite{Lynch_98-1}.

The arrival of an acoustic strain pulse to within a skin depth of
probe light results in a time-dependent strain profile
$\eta_{33}(z,t)$ which induces a change in the complex reflection
coefficient \cite{Thomsen_PRB_86-1,Saito_PRB_03-1}:
\begin{equation}
\frac{\delta \tilde{r}(t)}{\tilde{r}} =
-i\frac{4\pi}{\lambda}\int_0^{\infty}\Big(1-\frac{2\tilde{n}}{1-\tilde{n}^2}\frac{d\tilde{n}}{d\eta}{\rm
exp}\Big[i\frac{4\pi}{\lambda}\tilde{n}z\Big]\Big)\eta_{33}(z,t)dz\\,
\label{delta_r_over_r}
\end{equation}
with the complex refractive index $\tilde{n}=n+i\kappa$ and the
photoelastic coefficient
$d\tilde{n}/d\eta=dn/d\eta+id\kappa/d\eta$. Taking into account
the reflection of the acoustic pulse $\eta(t)$ from the gold-air
interface according to $\eta_{33}(z,t)=\eta(t+z/c_{\rm
s})-\eta(t-z/c_{\rm s})$, we can rewrite this expression in the
time domain
\begin{equation}
\frac{\delta \tilde{r}(t)}{\tilde{r}} = -i\frac{4\pi
v}{\lambda}\int_{-\infty}^{\infty}[1-(A+iB){\rm
e}^{i4\pi\tilde{n}v|t^{\prime}-t|/\lambda}]\sign(t^{\prime}-t)\eta(t^{\prime})dt^{\prime}\\.
\label{delta_r_over_r_time_domain}
\end{equation}
where the complex dimensionless parameter
$A+iB=2\tilde{n}(d\tilde{n}/d\eta)/(1-\tilde{n}^2)$ depends on
both the photo-elastic coefficient $d\tilde{n}/d\eta$ and the
index of refraction $\tilde{n}$.

The real part of this expression gives the reflectivity change
$S(t)$ measured in our pump-probe experiment:
\begin{equation}
S(t)=\frac{\delta R(t)}{R}=2{\rm Re}\Big[\frac{\delta
\tilde{r}(t)}{\tilde{r}}\Big]=\frac{8\pi
v}{\lambda}\int_{-\infty}^{\infty}G(t^{\prime}-t)
\eta(t^{\prime})dt^{\prime}\\, \label{S-function}
\end{equation}
with
\begin{equation}
G(t)=[A\sin(\omega_{\rm Br}t)+B\cos(\omega_{\rm Br}t)\sign
(t)]{\rm e}^{-|t|/\tau_{\rm skin}}\\. \label{G-function}
\end{equation}
Here we have introduced the Brillouin frequency\cite{Lin_JAP_1991-1} $\omega_{\rm
Br}=4\pi n v/\lambda=2\pi\times 29$~GHz and
the acoustic Time-of-flight $\tau_{\rm skin}=\delta_{\rm
skin}/v=4.8$~ps through the skin depth $\delta_{\rm
skin}=\lambda/(4\pi\kappa)=16.5$~nm (the complex index of
refraction of gold is $\tilde{n}=1.67+1.94i$ \cite{Lynch_98-1}  at
probe wavelength $\lambda=393$~nm).

The response function $G(t)$, which describes the
reflectivity variation induced by an infinitely short acoustic
pulse, represents the sum of two contributions.  The first sine-term
in Eq.~(\ref{G-function}) is a smooth function whose duration
$\tau_{\rm skin}$ is determined by the skin depth, whereas the
second cosine-term contains a sharp step. Fig.~2 shows both terms
of the response function, as well as examples of a total response
corresponding to $B/A = 1/2$ and $B/A = -1/2$. One can see
that the shape of the transient reflectivity response is
controlled by the ratio $B/A$:
\begin{equation}
\frac{B}{A} = \frac{\beta - \chi}{\beta\chi + 1}\,\,,\,\,\ \chi =
\Big(\frac{dn}{d\eta}/\frac{d\kappa}{d\eta}\Big)\,\,,\,\,\beta  =
\left( \frac{n}{\kappa} \right) \frac{n^2 + \kappa^2 -1}{n^2 +
\kappa^2+1}\,.
\end{equation}

\begin{figure}
\includegraphics[width=8 cm]{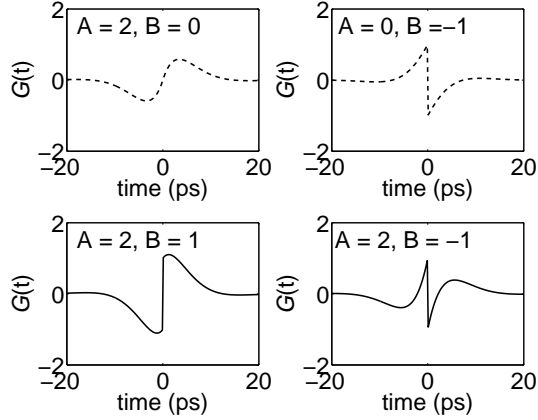}
\caption{\label{fig2} The response function $G(t)$ (see
Eq.~\ref{G-function}) for different combinations of the coefficients
$A$ and $B$.}
\end{figure}

It is instructive to analyze the time derivative $dS(t)/dt$ of the
transient reflectivity signal $S(t)$:
\begin{equation}
\frac{dS}{dt} =\frac{8\pi v}{\lambda}\left(2B\eta(t) +
\int_{-\infty}^{\infty}F(t^{\prime}-t)\eta(t^{\prime})dt^{\prime}\right)
\\, \label{derivative}
\end{equation}
with
\begin{equation}
F(t) =\left[\left(A\omega_{\rm Br}-\frac{B}{\tau_{\rm
skin}}\right)\cos(\omega_{\rm Br}t)- \left(B\omega_{\rm
Br}+\frac{A}{\tau_{\rm skin}}\right)\sin(\omega_{\rm
Br}|t|)\right]{\rm e}^{-|t|/\tau_{\rm skin}}
\\. \label{F-function}
\end{equation}
For a short acoustic pulse with duration $\tau_{\rm ac}<<\tau_{\rm
skin}$, the slowly varying integral term in Eq.~(\ref{derivative})
is proportional to $\tau_{\rm ac}$. Consequently, in the
short-pulse limit the first term is dominant and the time
derivative of the reflectivity response yields the acoustic strain
profile. Indeed, in the signal derivative shown in the inset in
Fig. 1(b), one can see a unipolar pulse of 2ps in duration.
However, the pulse is not short enough for the slow component of
the signal derivative to become negligible.

We see that in order to detect a short acoustic pulse, one needs a
material and probe wavelength combination that yields a large
value of $B$. One can see that $B$ vanishes for a weakly absorbing
medium with vanishing $\kappa$, hence a strongly absorbing material is
desired. On the other hand, B also vanishes in the limit $\kappa >> n$.
Thus a smaller skin depth is not necessarily preferred. Because
detailed data on the wavelength dependence of the photoelastic
constant $d \tilde{n} /d\eta$ are typically lacking, one has to resort to
trial-and-error in search for the best material / wavelength
combination.

We obtained an estimate of $B/A$ for gold by fitting
Eq.~\ref{derivative} to the time derivative of the reflectivity
using a simulated strain pulse. Because the low acoustic impedance
mismatch between STO and SRO suppresses the reflected wave
($Z_{\text{SRO}} = \rho_{\text{SRO}}v_{\text{SRO}}$ = 41$\times
10^6$ kg m$^{-2}$ s$^{-1}$  and $Z_{\text{STO}} =
\rho_{\text{STO}}v_{\text{STO}}$ = 40$\times 10^6$ kg m$^{-2}$
s$^{-1}$)\cite{Bell_PR_63-1}, we would expect the initial
rectangular strain pulse to have a FWHM~$\sim
d_{\text{SRO}}/v_{\text{SRO}}= 1.90$~ps. However, acoustic
dispersion and the difference in the arrival time at the gold-air
interface due to nanometer surface roughness cause the initial
rectangular pulse shape to spread\cite{Temnov_NatComm_13-1}.
Therefore, we assumed a Gaussian pulse shape and allowed the width
to vary to account for the overall acoustic broadening. The fit
parameters were the ratio $B/A$ and the FWHM of the Gaussian
profile. To improve the efficacy of the fit, "echoes" caused by
multiple reflections of the pump pulse inside the STO substrate
were removed by subtracting time-delayed and scaled signal from
the original data. The data pictured in Fig. \ref{fig1} is an
average over 25 scans; to test the repeatability of our fits, we
divided the raw data files into five sets of five scans each and
fit each set independently.

\begin{figure}
\includegraphics[width=8 cm]{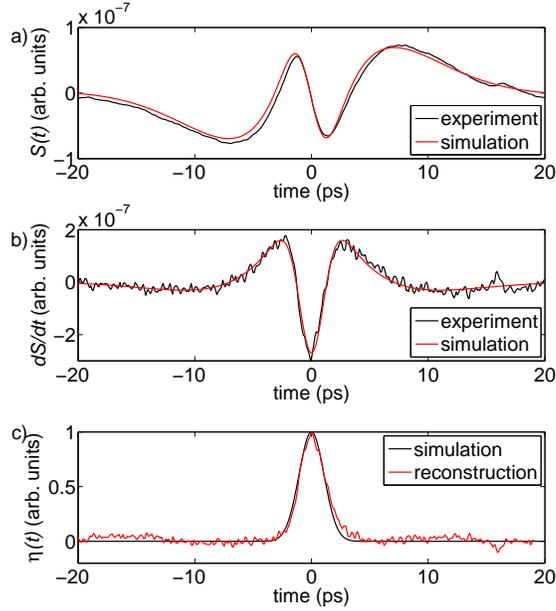}
\caption{\label{fig3} (a) A comparison of the simulated
reflectivity curve and the experimental data; (b) The
time-derivative of the reflectivity data and the fit that was
obtained for Eq.~\ref{derivative}; and (c) A comparison of the
simulated strain pulse and the reconstructed strain pulse.}
\end{figure}

The results of fitting the full data set are shown in
Fig.~\ref{fig3}(a-c), along with the simulated strain pulse. The
best results were obtained for $B/A =-0.59$ and FWHM $=2.54$~ps.
The statistical fitting error for $B/A$ was 0.01; however, based
on measurements of other samples we would suggest a more
conservative error estimate of 0.03. Using our fit value for
$B/A$, we obtained a ratio of $\chi$=1.9 at a wavelength of
393~nm. We expect a significant variation in the photoelastic
constants of gold at visible wavelengths due to an interband
transition in this region\cite{Devos_PRL_01-1}. The known
photo-elastic coefficients of other metals such as nickel and
chromium have the same order-of-magnitude ratio $|\chi|\sim
1$\cite{Saito_PRB_03-1}.

Quantitative acoustic measurements require a reliable algorithm
for the reconstruction of an acoustic pulse profile $\eta(t)$. It
was demonstrated recently that for ultrafast acousto-plasmonic
measurements, the strain profile can be recovered from the
plasmonic analog of Eq.~(\ref{derivative}) using an iterative
numerical method \cite{Temnov_NatComm_13-1}. However, in the case
of reflectivity measurements, the coefficients of the sine and
cosine terms in Eq.~(\ref{F-function}) are too large and the
iterative solution of Eq.~(\ref{derivative}) diverges.

A more general method for pulse reconstruction is demonstrated
here. After taking the Fourier transform of Eq.~\ref{S-function}
and applying the convolution theorem, the reconstructed strain in
the Fourier domain is given by $\eta(\omega)=(\frac{\lambda}{4\pi
v})S(\omega)/G(\omega)$ with
\begin{equation}
G(\omega)=\int_{-\infty}^{\infty}G(t) {\rm e}^{-i\omega t}dt=
-2i\omega\frac{(2\omega_{\rm Br}/\tau_{\rm
skin})A+(\omega^2-\omega_{\rm Br}^2+1/\tau_{\rm
skin}^2)B}{(\omega^2-\omega_{\rm Br}^2+1/\tau_{\rm
skin}^2)^2+(2\omega_{Br}/\tau_{\rm skin})^2} \label{G-function-omega}.\
\end{equation}
Applying the inverse Fourier transform delivers the desired
acoustic strain profile \begin{equation}
\eta(t)=\frac{\lambda}{8\pi^2v}\int_{-\infty}^{\infty}\frac{S(\omega)}{G(\omega)}
{\rm e}^{i\omega t}d\omega\\, \label{reconstructed_eta}
\end{equation}
shown in Fig.~3(b). The reconstructed waveform matches the
Gaussian pulse obtained by curve fitting fairly well. However, the
reconstruction algorithm did not use any assumptions about the
acoustic pulse shape and can be employed to recover arbitrary
acoustic strain profiles.

The reconstructed waveform contains barely visible slow
oscillations of the baseline which become more prominent when
noisier data are analyzed. The spectral function $G(\omega)$
possesses three zeros, $\omega=0$ and $\omega=\pm\omega_0=\pm
2\pi\times 56$~GHz, resulting in singularities in
$\eta(\omega)=(\frac{\lambda}{4\pi v})S(\omega)/G(\omega)$ in the
presence of noise with non-zero spectral components at these
frequencies. The singularity at $\omega=0$ is prevented by
subtracting the slow background from the experimental signal
$S(t)$ to guarantee $\int S(t)dt=0$ ($S(\omega=0)=0$). The two
other roots $\omega=\pm\omega_0$ result in a background modulation
with a period of $2\pi/\omega_0=18$~ps. This modulation does not
present a problem for the analysis of shorter-than-skin-depth
acoustic pulses but may become a nuisance for longer waveforms
containing significant frequency components at $\omega_0$. The
existence of non-zero roots of $G(\omega)$ requires $\omega_{Br}^2
- 1/\tau_{skin}^2 > (A/B)(2\omega_{Br}/\tau_{skin})$ and can be
avoided if  different probe wavelengths and/or a different metal
are used.

In conclusion, our results demonstrate that transient reflectivity can be used to
detect shorter-than-skin depth acoustic pulses in common metals
such as gold. The ability to resolve short acoustic pulses depends
is not limited by the the skin depth of the material; rather, it
is controlled by the parameter B/A that depends on the
photoelastic constants and the  the complex index of refraction at
the detection wavelength. We have demonstrated a pulse
reconstruction algorithm capable of recovering the acoustic pulse
profile from transient reflectivity data which will advance
quantitative ultrafast acoustic spectroscopy.

This work was supported in part by U. S. Department of Energy
Grant DE-FG0200ER15087 and {\it Nouvelle \'{e}quipe, nouvelle
th\'{e}matique de la R\'{e}gion Pays de La Loire}. The work of at
University of Wisconsin-Madison was supported by the AFOSR through
grant FA9550-12-1-0342 and Army Research Office through grant
W911NF-10-1-0362. K.J.M. acknowledges support from a National
Science Foundation Graduate Research Fellowship. D.M. acknowledges
financial support from the European Research Council under the
European Union's Seventh Framework Programme (FP7/2007-2013) / ERC
grant agreement number 306277.

%

\end{document}